\documentstyle[12pt,aaspp]{article}

\begin{document}

\newcommand{\cosb}{\hbox{\em COS-B}}
\newcommand{\sas}{\hbox{\em SAS-2}}
\newcommand{\egret}{\hbox{\em EGRET}}

\newcommand{\Int}{\displaystyle \int}
\newcommand{\bul}{\hbox{$\bullet$}}
\newcommand{\PP}{{\hbox{$\gamma\gamma \rightarrow e^+e^-$}}}
\newcommand{\cdt}{{\hbox{~$\cdots$~}}}
\newcommand{\dagg}{{\hbox{$^\dagger$}}}

\newcommand{\E}{\varepsilon}
\newcommand\vvmax{\hbox{$\langle V/V_{\rm max}\rangle$}}
\newcommand\vvmaxp{\hbox{$\langle V^\prime/V^\prime_{\Max}\rangle$}}
\newcommand\veva{\hbox{$\langle V_e/V_a \rangle$}}
\newcommand\vmax{\hbox{$V_{\rm max}$}}
\newcommand\nlim{\hbox{$n_{\rm lim}$}}

\newcommand{\fsyn}{{\rm F}_{\rm syn}}
\newcommand{\Psf}{{\rm psf}}
\newcommand{\PSF}{{\rm PSF}}
\newcommand{\sar}{{\rm SAR}}
\newcommand{\edp}{{\rm EDP}}
\newcommand{\Like}{{\cal L}}
\newcommand{\TS}{{\rm TS}}
\newcommand{\Max}{{\rm max}}
\newcommand{\Min}{{\rm min}}

\title{An Outer Gap Model of High-Energy Emission from
Rotation-Powered Pulsars}
\author{James Chiang\altaffilmark{1} and Roger W. Romani\altaffilmark{2}}
\affil{Department of Physics, Stanford University, Stanford CA
94305-4060}
\altaffiltext{1}{Present Address: CITA, 60 St. George St., Toronto,
Canada, M5S 1A1}
\altaffiltext{2}{Alfred P. Sloan Foundation Fellow}
\dates

\begin{abstract}
We describe a refined calculation of high energy emission from
rotation-powered pulsars based on the Outer Gap model of Cheng, Ho
\&~Ruderman (1986a,b). In this calculation,
vacuum gaps form in regions near the speed-of-light
cylinder of the pulsar magnetosphere along the boundary between the
closed and open field line zones. We have improved upon previous efforts
to model the spectra from these pulsars (e. g. Cheng, et al. 1986b; Ho 1989)
by following the variation in particle production and radiation
properties with position in the outer gap. Curvature, synchrotron
and inverse-Compton scattering fluxes vary significantly over the
gap and their interactions {\it via} photon-photon pair production
build up the radiating charge populations at varying rates.
We have also incorporated an approximate treatment of the transport of
particle and photon fluxes between gap emission zones. These effects,
along with improved computations of the particle and photon
distributions, provide very important modifications of the model
gamma-ray flux. In particular, we attempt to make specific predictions
of pulse profile shapes and spectral variations as a function of pulse phase
and suggest further extensions to the model which may provide
accurate computations of the observed high energy emissions.
\end{abstract}

\section{Introduction}

Since the discovery of radio pulsations from rotating neutron stars
(Hewish et al.~1968), many theories have been proposed to explain the
origin of these emissions. Early models invoke the acceleration of
electrons and positrons to extremely high energies such that the
radiation emitted by these particles results in an electromagnetic
cascade of $e^\pm$ pairs and photons. This avalanche of charges yields
the observed coherent radio flux (Sturrock 1971; Ruderman \&
Sutherland 1975). Unfortunately, attempts at understanding the physics
of the pulsar magnetosphere by looking only at the radio observations
have met with limited success.  A more fruitful approach would be to
use the higher energy observations, both in terms of photon energy and
in terms of the total available energy to be extracted from the
pulsar, to characterize the magnetosphere (Arons 1992).  The polar cap
models of Harding, Tademaru, \&~Esposito (1978), Daugherty \&~Harding
(1982), the slot gap model of Arons and his collaborators
(Scharlemann, Arons, \&~Fawley 1978; Arons 1983)
and the outer gap model of Cheng, et al. (1986a,b; hereafter CHRa,b)
and Ho (1989) have all been attempts to use the high energy
observations to reveal nature of the pulsar magnetosphere.

Given the recent data obtained by the instruments aboard the {\it
Compton Gamma-Ray Observatory} and the identification of additional
gamma-ray pulsars, bringing the total to six---Crab, Vela, Geminga
(Bertsch et al. 1992), PSRB~1706$-$44 (Thompson et al. 1992),
PSRB~1055$-$52 (Fierro et al. 1993) and PSRB~1509$-$58 (Wilson et
al. 1992)---it is now appropriate for further efforts to be made
towards understanding these objects.  In this paper, we present our
own attempts at modeling the high energy emission from young,
rotation-powered pulsars in the context of a modified outer gap
model. In section~2, we summarize the results of a previous paper
(Chiang \& Romani~1992; hereafter Paper~I), in which a geometrical
calculation of the emission from the outer gap regions was performed
and which was shown to reproduce qualitatively the light curves of all
the gamma-ray pulsars. In section~3, we review the details of the
outer gap spectral calculation described in CHRa,b and restated later
by Ho (1989); and in section~4, we motivate and outline our
refinements to the CHR calculation. They consist of dividing the outer
gap region into smaller sub-zones in order to account for the
variation in magnetic field and photon and particle densities as a
function of position in the magnetosphere. We have also included a
treatment of the transport of particles and photons from one region of
the magnetosphere to another.  We find that this transport crucially
effects the emission processes in the outer gap.  In addition, we
perform more detailed radiation and pair creation calculations. In
section~5, we present some preliminary results of our multi-zone
spectral calculations; and in section~6, we discuss limitations of the
model and ways in which it can be further improved.

\section{A Simple Model for the Outer Gap Light Curves}

In Paper~I, we presented our basic geometrical model of emission
from the outer gap regions of a pulsar magnetosphere. Adopting the
assumptions of CHRa, we defined the outer gap to lie along the
boundary of the closed field line region of a magnetic dipole field
configuration, bounded on the side closest to the neutron star by the
null-charge surface, on which ${\boldmath\Omega} \cdot {\boldmath B} =
0$, and on the outside by the velocity-of-light cylinder. However, in
contrast to CHRa, we assumed that gap-type regions could be supported
along all field lines on the boundary of the closed region rather than
just on the bundle of field lines lying in the plane of the rotation
and magnetic dipole axes.

	Assuming uniform emissivity along outer gap fields lines with
the radiation beamed in the local field direction, we then computed
the emission profile of the pulsar projected onto the sky. In this
model, the field geometry is that of a dipole in the co-rotating
frame, and we have included the effects of relativistic aberration and
time-of-flight delays through the magnetosphere.
Figure~\ref{OuterGap_image} shows the result of our calculation for a
nearly orthogonal rotator. As in Paper~I, the coordinates of this
image are defined so that rotational latitude runs along the vertical
axis and rotational longitude runs along the horizontal axis. Hence,
an observer line-of-sight corresponds to a horizontal line across the
image, and the zero of rotational phase is defined to be where the
line-of-sight crosses the plane containing the dipole and rotation
axes. A line-of-sight which has a phase separation of $\sim 145\deg$
between the two highest peaks in the emission profile is indicated by
the dashed line in the image. A combination of the time-of-flight and
relativistic aberration calculations allows the observer to see
emission from many points along the outer gap at once during certain
portions of the pulse phase.  Phases where this effect is strongest
correspond to the enhancements or peaks of the emission.

Given this emission profile on the sky, it is apparent that the
various pulse shapes seen for the gamma-ray pulsars can be accounted
for {\em generically} by this model. For almost all lines-of-sight,
two main peaks with significant emission in the ``bridge'' region are
seen.  Thus, pulse shapes such as those of the Crab, Vela and Geminga
pulsars are seen if the observer line-of-sight were at rotational
latitudes $\la 20\deg$, whereas the light curves for PSRB~1706$-$44
and PSRB~1055$-$52 correspond to lines-of-sight at higher latitudes
near $\sim 40\deg$ where the separation of the two peaks is small and
may appear as a single, broad peak.  Furthermore, this model does not
require particularly special alignment of the dipole axis to the
rotation axis.  So long as the angle, $\alpha$, between these two axes
is relatively oblique (i.e.  $\alpha \ga 45\deg$), emission similar to
that pictured in Figure~\ref{OuterGap_image} will result. The cases
where pulsed emission is unlikely to be observed in this model occur
when the axes are nearly aligned. In the extreme case of $\alpha =
0\deg$, the emission is, by symmetry, completely unmodulated.

	Additionally, because the aberration and time-of-flight
effects make it possible to attain the double pulse morphology for
emission from a single hemisphere of the pulsar dipole, double pulses
with bridge emission are present for more general field configurations
than just the neutron star-centered dipole models as invoked by
traditional polar cap models and by the outer gap model proposed in
CHRa to explain the Crab and Vela pulse profiles. Our geometrical
outer gap model also constrains the phase of emission from the polar
cap relative to the gamma-ray emission. Either radio or thermal
emission, both believed to originate from the polar cap regions, will
lead the first (or single) peak of the gamma-ray light curves
predicted by our model. Calculations show that the model matches
quantitatively the relative phases observed for the individual
gamma-ray pulsars as well as predicting relative numbers of detections
in good agreement with the present sample (Romani
\&~Yadigaroglu~1993). In addition, Romani \&~Yadigaroglu (1993) have
shown that the position angle variation of the optical polarization of
the Crab pulsar (Kristian et al.~1970; Smith et al.~1988) is well
matched to predictions of our outer gap model. With these results, the
basic geometrical
picture is well established and we turn now to addressing the
the observed spectral variations as a function of pulse phase. These
variations, absent in the original outer gap picture, arise as a natural
consequence of our geometrical model, but require a careful calculation
of {\it local} emission properties.

\section{Review of the Original Cheng, Ho and Ruderman Calculation}

Before proceeding with a description of our spectral calculations, it
is appropriate to review the basic outer gap spectral calculation
which was presented by CHRa,b.  According to CHRa, the outer gap
is a vacuum zone which is stable against closure by photon-photon
pair-production because of the special geometry of the magnetic field
at the boundary of the closed field line region. The deviation from
co-rotation charge density in this region results in large potential
drops on paths parallel to the local magnetic field. Electrons and
positrons produced near the pair-creation boundary of the gap are
accelerated to have Lorentz factors of $\Gamma \sim 10^7$.  Their
energies are radiation-reaction limited by a combination of curvature
radiation and inverse-Compton processes. Because the electrons and
positrons are highly relativistic, the radiation they emit is
beamed along the local field direction.  The low energy photons
off which the high energy particles inverse-Compton scatter are
produced by counter-streaming electron and positron pairs which are
created just beyond the outer gap pair boundary and whose momentum
vectors can form sizable angles with respect to the local field
direction. These secondary pairs emit synchrotron radiation and also
inverse-Compton scatter both ambient and low-energy photons produced
from secondary particles traveling in the opposite direction. Because
the secondary pairs are created outside the vacuum gap, they do not
attain large Lorentz factors and the radiation they emit
has greater angular dispersion than that of the highly
relativistic primaries. This allows a fraction of the secondary
radiation to be projected into the gap zone and forms the
pair-production opacity which limits the size of the gap.  The gap
size is self-consistently controlled by the pair-production of the
counter-streaming primary and secondary photon beams, and the
radiation processes are themselves maintained via the pair-production
processes which provide the radiating charges.

One of the major deficiencies of the outer gap spectral calculations
of CHRa and Ho (1989) is that they consider the outer gap as a
single, monolithic zone with only single, ``characteristic'' values of
the magnetic field and photon densities which do not vary in
space. Furthermore, because of the built-in symmetry of the CHR
spectral calculation (see Ho 1989), the spectra for the two peaks
(which corresponds to the inward and outward radiation beams) are
identical. Therefore, in addition to being unable to account for the
emission in the bridge region between the main pulses of the Crab,
Vela and Geminga light curves as well as the single broad pulses seen
in the light curves of PSRB~1706$-$44 and PSRB~1055$-$52, the outer
gap model of CHR is also unable to account for the spectral variation
seen as a function of phase for the double-peaked gamma-ray pulsars.
In our model, the variation in the magnetic field and photon densities
will have a significant effect on the resulting radiation and thus
provides a natural explanation of the observed spectral variation as a
function of pulse phase.

\section{Refinements}

\subsection{Mapping pulse phase to position in the magnetosphere}

Figure~\ref{coarse_sampling} shows the same outer gap emission
calculation as does Figure~\ref{OuterGap_image}, but with the emission
from along just twenty field lines traced from null surface to the
light cylinder. This coarser sampling of field lines allows us to
infer the approximate location of the emission at each point in phase
along a given line-of-sight. Emission points along the field lines are
drawn so that the density of points is the same for each line.
Therefore, the number of emission points along a given field line
from the null surface out to a specified point on the skymap image
will be proportional to the path length from the null surface to that
point and will correspond to a specific location in the magnetosphere.

Along the line-of-sight which has the two highest peaks of the
emission separated by a Crab- or Vela-like phase difference
($\Delta\phi \simeq 145\deg$), we see from
Figure~\ref{coarse_sampling} that the first peak is composed of
radiation which comes from fairly high in the gap region and the
second peak is composed of photons originating from almost the entire
length of a bundle of field lines. By contrast, the emission in the
bridge region appears to consist of photons from regions very near the
light cylinder and regions very near the null surface but not much
from between the two regions.

\subsection{The Multi-zone Outer Gap}

Given that the pulsar spectra vary in phase and that the phases of the
emission of the light curves can be mapped back to different locations
in the magnetosphere, it is natural to model the spectral
emissivity locally and divide the magnetosphere into smaller sub-zones
so that each zone can be treated as a separate emission region with
a distinct magnetic field intensity and geometry.
Figure~\ref{gap_sub-zones} is an example of how we have divided the
outer gap into smaller sub-zones in the plane of the rotation and
dipole axes.  The photon densities will also be different for each
zone, and as we shall see, will differ for the two beaming
directions.  Since the observed pulsar radiation is mostly beamed
along the local field direction\footnote{For the Crab pulsar, we know
that the observed radiation must include a sharply beamed component
from optical to gamma-ray
energies since the light curves in all these energy bands are sharply
peaked at the same points in pulse phase.}, we model the gap as a
one-dimensional region existing along the last closed field lines,
radiate photons along paths initially
parallel to the local field direction and ignore the contribution of
neighboring zones which are displaced toroidally.

Along with the variation of the magnetic field, which can range from
$B \sim 10^{11}$~G at the null surface for a nearly orthogonally
aligned Crab-like pulsar to $B \sim 10^{6}$~G in regions near the
light cylinder, another important motivation to subdivide the gap
region is the possibility of photon and particle transport from one
part of the magnetosphere to another. Photon transport is
important for two reasons.  First, at low energies, the optical depth
to photon-photon pair-production is small, meaning that the low energy
photons will be able to travel freely from sub-zone to sub-zone.
Since the low energy photons contribute most to the
pair-creation opacity seen by the high energy energy photons, it is
crucial to include their transport.  Figure~\ref{optical_depths} shows
the optical depth to photon-photon pair-production as a function of
energy at three points along the gap boundary.  Second, assuming that
the boundaries of the gap zone follow dipole field lines which are
taken to be equipotentials (see CHRa and the next section), the
geometry of the outer gap is such that the flux contribution to a
given sub-zone from sub-zones farther out in the gap will
not be equal to the contribution from those closer to the null
surface (see Figure~\ref{acceptance_angles}).  This asymmetry causes
the pair-production rate to differ for the inward and outward
directions and ultimately leads to a difference in observed photon
flux.

The importance of the particle transport can be estimated by comparing
the energy lost due to synchrotron emission and inverse-Compton
scattering to the total initial energy of the particle.  In
Figure~\ref{particle_energy_losses}, we have plotted the fraction
of energy remaining to each of the secondary electrons and positrons
as they cross single sub-zones versus the initial energy of each particle.
It is evident that a substantial number of electrons retain a
significant fraction of their initial energy after they cross a single
sub-zone.

\subsection{Other Improvements and Approximations}

In addition to dividing the gap into multiple zones and a crude
treatment of particle and radiation transport, we have
improved upon the methods used to calculate the inverse-Compton photon
distributions as well as the electron-positron distributions due to
photon-photon pair-production. In CHRb and Ho (1989), very generous
approximations were made to calculate these processes and in some
cases were not entirely justified. We have been able to calculate
semi-analytically the pair and inverse-Compton photon distributions
without making the same simplifying assumptions of the previous
works.

{}From the pair-production calculations, we use the angular and energy
distribution information of the resulting secondary particles to
determine the synchrotron radiation angular distribution in each
sub-zone as a function of photon energy. This information is then used
with the sub-zone photon acceptance angles (see
Figure~\ref{acceptance_angles}) to determine the amount of radiation
passed from sub-zone to sub-zone in the photon transport part of the
calculation.  However, when determining the distributions of locally
generated radiation and particles, we still adopt the approximation
that the incident particles and photons in the pair-creation and the
inverse-Compton processes impinge upon each other {\em head-on}.  This
assumption is somewhat justified for the photon/photon interactions by
the fact that the observed radiation (at least in the case of the
Crab) must be beamed.

Another approximation we make, and which limits the predictive power
of this model, is our prescription for determining the upper boundary
of the outer gap.  If we assume that the field lines in the open
region are equipotentials and that charges can flow freely along them,
the upper boundary of the gap will lie along these field lines.  This
is implied by our depiction of the gap sub-zones in
Figure~\ref{gap_sub-zones}.  Ho (1989) chose to parametrize the width
of the gap as a fraction of the radius of curvature of the local
magnetic field: $f_g \equiv a_{\rm gap}/r_{\rm curv}$. This is a
natural parametrization since the rectilinear approximation of the gap
potential (CHRb) yields a voltage and a current density through the
gap---both of which can be expressed in terms of the parameter $f_g$.
For this paper, which we restrict to the case of the Crab
pulsar, we chose a value of $f_g = 0.3$, evaluated at the midpoint of
the gap, which yields a value of the gap power which is consistent
with the observed pulsed radiative power emitted by the Crab, and
which is still far less than the total spindown power.

\section{Spectral Results}

Obtaining the final spectra for all the gap sub-zones requires an
iterative calculation to be performed so that the pair-production
rates, which depend on the photon-photon optical depths, and the
photon fluxes, which depend on the charged particle distributions, are
consistent throughout the gap. Ho (1989) presents an example of an
iterative, single-zone spectral determination.  Our calculation is
similar, but entails additional complications due to the transport of
radiation and particles from sub-zone to sub-zone.

Following Ho (1989), we use a Crab-like spectrum, taken from the
observations, to start off our iterative calculations. The result of
the initial iteration of the calculation is shown in
Figure~\ref{initial_iteration}. The spectra from three points along
the gap are shown. The dashed line is the input Crab-like spectrum,
and the solid line is the summed flux emitted from each zone. For this
initial iteration, the results look promising: In the first
sub-zone, which is nearest the star and thus has the strongest magnetic
field, the synchrotron contributions dominate the emission and have a
characteristically steeper spectrum with photon spectral index $\gamma
\sim 2$. For the middle and outermost zones, the spectra are progressively
flatter and reflect the increased importance of the inverse-Compton
processes in the outer magnetosphere where the fields are weaker and
the low energy flux is greater because of the larger photon acceptance
angles (see Figure~\ref{acceptance_angles}).
Figure~\ref{flux_components} shows the relative contribution of the
various spectral components to the total spectrum for the innermost
and outermost gap zones.

Unfortunately, once the calculation is allowed to converge to a
self-consistent solution, the spectra which result exhibit a
significant lack of photon flux below several GeV. The expected power
is still being extracted from the gap, except that it is emerging in
the form of very high energy photons at energies well above a GeV.

\section{Discussion}

The root of the problem may lie in several different places.  Since
the gap power is being extracted as very high energy photons, the
solution amounts to reprocessing this flux to lower energies. A more
traditional pair-photon cascade would be an obvious means of doing
this. There is apparently too little optical depth to pair-creation
from just the low energy photons created by the counter-streaming
particles themselves, and the magnetic fields are too weak in the
outer magnetosphere to contribute to the pair-creation via the
Sturrock process, $\gamma B \rightarrow \gamma\gamma$ (Sturrock~1971).

External sources of soft flux, from processes other than the
ones described here, may be sufficient to maintain the needed
pair-production.  One source may be thermal emission originating from
polar cap heating by inwardly streaming particles. Recent ROSAT
observations of Geminga and PSRB 1055-52 indicate that there is pulsed
thermal emission from these pulsars which is consistent with local
heating on the surface of the neutron star (Halpern \& Holt 1992;
\"Ogelman 1993). However, this emission would most likely affect only
the inner portion of the magnetosphere.

A more complicated solution may lie in the manner in which the low
energy flux is transported between different regions of the
magnetosphere. In the present calculation, the soft photon fluxes are
only transported along the gap in the poloidal direction. For the
beamed, high energy component of the emission, this approximation is
probably valid.  However, at lower photon energies ($\E \la 1$~keV),
the beaming is not as strong, particularly if the synchrotron
component dominates this part of the spectrum. Therefore, for lower
photon energies where the bulk of the radiation is emitted into large
angles, the emission from each zone should be treated as if it were
almost isotropic rather than beamed along the field lines.  Isotropic
treatment of the low energy emission may make possible the transport
of soft flux from the inner magnetosphere to regions in the outer
magnetosphere where it is needed to produce pairs which would in turn
produce more secondary radiation, the higher energy component of which
would be beamed along the local magnetic field.  The difficulty of
implementing a calculation with three-dimensional transport is that it
requires convergence of the generated spectra at each iteration over
{\em all} the interacting zones of the magnetosphere at once rather
than the relatively simpler task of convergence among the only $\sim
20$ zones at a time in the present calculation.

Another solution may come from a more realistic calculation of the
secondary electron distribution function.  Throughout our calculation
we have assumed that interactions are entirely ``head-on''.  This
assumption makes the form of the pair-production computations
sufficiently simple so that they can be performed semi-analytically.
However, if we relax this constraint, we see that the threshold for
pair-production varies significantly as a function of incident
angle. For photons with an angle $\eta$ between their momentum vectors
and lab frame energies $\E_1$ and $\E_2$, the threshold condition for
pair-production is
$$
\E_1\E_2 \ge {2m_e^2\over 1 - \cos\eta}.
$$
This implies that larger photon energies are required
in the lab frame to produce pairs for smaller values of $\eta$,
and since the mean energy of the secondary electrons in the lab frame is
$\sqrt{\E_1\E_2}$, smaller values of $\eta$ will result in higher mean
energies of the created pairs and harder secondary electron
distributions.

For a source of photons at the surface of the star such as a hot spot
at the polar cap, the inwardly streaming primary photons will see
incident angles which are more nearly head-on ($\eta \sim 180^\circ$),
whereas the outwardly streaming primary photons will see smaller
values of $\eta$.  Thus we expect that the inwardly going secondary
electrons will have a softer spectrum than the outward ones. This
steeper spectrum for the inward secondary electrons could provide the
enough low energy photons to enhance the radiation produced by
inverse-Compton scattering by the outward flowing electrons, and thus
bootstrap the spectrum to produce the observed Crab-like spectrum in
the outward beam.

\section{Conclusions}

We have described our efforts to refine the standard outer gap picture
of high energy emission from pulsar magnetospheres. Our light curve
calculations, in our more general outer gap geometry, are able to
account for the various pulsar profiles seen for the gamma-ray pulsars
observed by the instruments aboard the {\em Compton Observatory}.
These light curve calculations also point to the source of the
observed spectral variation with phase seen for the gamma-ray pulsars:
there is a clear mapping of location in the magnetosphere to pulse
phase. We have thus attempted to carry through the outer gap spectral
calculation as outlined by CHRb and Ho (1989) in our modified
geometry, including the refinements of radiation and particle
transport and improved inverse-Compton and pair-production
calculations. Our efforts have met with limited success. The
calculation suffers from too little low energy photon flux and the
converged pulsar spectra we produce are too hard for all phases of the
pulse profile.  However, our results indicate that a realistic
computation of the pulsar emission requires at least an approximate
three-dimensional treatment of the radiation transport in the outer
magnetosphere. Since photon fluxes themselves provide the relevant
opacities, this is a computationally difficult, non-linear
problem. However, the spectral differences between the competing
radiation processes illustrated in our sample calculations do point
the way for a more satisfactory description of the high-quality pulsar
data from the {\em Compton Gamma-Ray Observatory}.
\bigskip
\bigskip

	We thank Cheng Ho for participating in important initial
discussions on the physics of gap zones and for sharing a copy of his
outer gap spectral code. RWR was supported in part by NASA grants
NAGW-2963 and NAG5-2037. JC was supported by NASA grant NAG5-1605 and
gratefully acknowledges useful discussions with {\em EGRET} team
members at Stanford University, Goddard Space Flight Center and the
Max-Planck-Institut f\"ur Extraterrestriche Physik at Garching,
Germany.

\clearpage

\begin{figure}
\caption{The upper panel is the uniform emission from the outer gap
projected onto the sky. The dashed line-of-sight corresponds to a
difference in phase between the main peaks of $\sim 145\deg$
appropriate for a Crab- or Vela-like phase difference. The
corresponding light curve is shown in the lower panel.}
\label{OuterGap_image}
\end{figure}

\begin{figure}
\caption{A coarse sampling of emission points in the outer gap. Field lines
with different azimuthal angles around the polar cap are followed from
the null charge line to the light cylinder in order to determine the
origin of emission as a function of pulse phase.}
\label{coarse_sampling}
\end{figure}

\begin{figure}
\caption{An example of the division of the outer gap along a field line
into smaller sub-zones.}
\label{gap_sub-zones}
\end{figure}

\begin{figure}
\caption{Photon optical depth for the gap sub-zone nearest the null charge
surface, the sub-zone in the middle of the gap, and the sub-zone nearest the
light cylinder (clockwise from upper left). The solid lines are the optical
depths seen by the outward traveling photons, and the dashed lines are the
optical depths seen by the inward traveling photons.}
\label{optical_depths}
\end{figure}

\begin{figure}
\caption{Photon acceptance angles between sub-zones along an outer gap field
line. Darker pixels indicate a larger acceptance angles.  The gap
sub-zones are numbered starting from the innermost sub-zone at the
null-charge line. The vertical axis labels the sub-zone from which the
locally generated radiation originates. The horizontal axis labels the
sub-zone into which the photons are transported.}
\label{acceptance_angles}
\end{figure}

\begin{figure}
\caption{The fraction of energy lost by particles flowing along the outer gap
as they traverse a single sub-zone. The vertical axis is the fraction
of energy lost, the horizontal axis is the original electron energy.
The darkness of the pixels represent the multiplicities in each bin.}
\label{particle_energy_losses}
\end{figure}

\begin{figure}
\caption{The spectra from the self-consistent outer gap calculation for the
initial iteration. Gap points 2, 7 and 19 correspond to inner, middle and
outer gap sub-zones, respectively (cf. Figure~3).}
\label{initial_iteration}
\end{figure}

\begin{figure}
\caption{Components of the outer gap spectral calculation for the initial
iteration.  The dashed line is the synchrotron component, the dotted
line is the primary radiation (curvature + inverse-Compton by the
primary electrons) and the dash-dotted line is the inverse-Compton
component from the secondary electrons. In the upper panel, the
spectra from a gap sub-zone near the null-surface are plotted; and in
the lower panel, the spectra from a gap sub-zone near the light cylinder
are shown.}
\label{flux_components}
\end{figure}

\end{document}